\documentclass[aps,prl,reprint,groupedaddress]{revtex4-1}

\usepackage{amsmath}
\usepackage{txfonts}

\usepackage[T1]{fontenc}
\usepackage{textcomp}
\usepackage[utf8]{inputenc}

\usepackage[pdftex]{graphicx}

\usepackage{enumerate}

\usepackage{enumerate}

\usepackage{bm}
\bmdefine{\aVector}{a}
\bmdefine{\AVector}{A}
\bmdefine{\bVector}{b}
\bmdefine{\BVector}{B}
\bmdefine{\cVector}{c}
\bmdefine{\CVector}{C}
\bmdefine{\eVector}{e}
\bmdefine{\EVector}{E}
\bmdefine{\fVector}{f}
\bmdefine{\FVector}{F}
\bmdefine{\gVector}{g}
\bmdefine{\pVector}{p}
\bmdefine{\PVector}{P}
\bmdefine{\qVector}{q}
\bmdefine{\QVector}{Q}
\bmdefine{\rVector}{r}
\bmdefine{\RVector}{R}
\bmdefine{\sVector}{s}
\bmdefine{\uVector}{u}
\bmdefine{\vVector}{v}
\bmdefine{\VVector}{V}
\bmdefine{\muVector}{\mu}
\bmdefine{\OmegaVector}{\Omega}

\begin{document}


\title{Thermodynamic stability of multicomponent ideal gas}

\author{Yuki Norizoe}
\author{Toshihiro Kawakatsu}
\affiliation{Department of Physics, Tohoku University, 980-8578 Sendai, Japan}




\date{April 28, 2018}

\begin{abstract}
We present an example counter to the widely-accepted concept on equilibrium states that ``Any thermodynamic equilibrium state of two component systems is determined by specifying 4 thermodynamic variables that include at least 1 extensive variable.'' which is equivalent to Gibbs' phase rule. We demonstrate this fact by thought experiments on an A/B binary mixture where temperature, pressure and chemical potential of the A-species are chosen as the 3 intensive variables while the number of B-molecules is chosen as the 1 extensive variable. Our findings also apply to $M (>2)$-component systems.
\end{abstract}


\maketitle

Here we show fundamental findings of binary thermodynamics in conventional ensembles, which we tentatively define as the ensembles with one or more external extensive variables, e.g. microcanonical ensemble and canonical ensemble. Thermodynamics~\cite{Atkins:PhysicalChemistry2018,Fermi:Thermodynamics,Greiner:ThermodynamicsAndStatisticalMechanics,Callen:ThermodynamicsAndAnIntroductionToThermostatistics} is a self-contained scientific field and forms the basis for statistical mechanics, quantum mechanics, and other fields. However, thermodynamic stability of binary systems in conventional ensembles was overlooked in previous works. We shed light on this problem and reveal equilibrium conditions and non-equilibrium phenomena of binary systems in the present work utilizing thought experiments. Our results shown in the present article will enhance a basic understanding of binary thermodynamics that is common and essential to studies on various physical phenomena of binary systems.

Arbitrary value sets of two external intensive and one external extensive variables of any single component conventional ensemble, e.g. (temperature $T$, number density of particles $\rho = N / V$, system volume $V$) of the canonical ensemble ($NVT$-ensemble) where $N$ denotes the number of particles, are independently and simultaneously given from the outside and always result in equilibrium. The two external intensive variables, e.g. $(T, \rho)$ of $NVT$-ensemble, determine the phase and state of the single component system. On the other hand, the one external extensive variable, e.g. $V$ of $NVT$-ensemble, sets the scale of the system. In other words, the phase and state of the system are dependent on the two external intensive variables and independent from the one external extensive variable. This indicates that thermodynamic degrees of freedom of any single component system equal two and results from Gibbs-Duhem equation,
\begin{equation}
	\label{eq:Gibbs-DuhemMulti}
	S \, dT - V \, dP + \sum_{\alpha} N_{\alpha} \, d\mu_{\alpha} = 0,
\end{equation}
where $S$ and $P$ denote the entropy and pressure, respectively. The suffix $\alpha$ represents the particle species. $N_{\alpha}$ and $\mu_{\alpha}$ denote the number of $\alpha$-particles and chemical potential of $\alpha$-species respectively. The summation $\sum_{\alpha}$ runs over the particle species. This result is consistent with Gibbs' phase rule. A straightforward extension of this result to binary systems suggests that arbitrary value sets of 3 external intensive and one external extensive variables of any conventional ensemble with 2 particle species, e.g. $( T, P, \mu_\text{A}, N_\text{B} )$ of $\mu_\text{A} N_\text{B} PT$-ensemble which is discussed below, can be independently and simultaneously given from the outside and always result in equilibrium. Thermodynamic degrees of freedom of the binary systems, which equal 3, are satisfied in any conventional ensemble. However, the results of our simple thought experiments shown below in the present work totally contradict this extension that has been considered correct at least since Gibbs-Duhem equation, eq.~\eqref{eq:Gibbs-DuhemMulti}, was found. The results indicate that the binary system in conventional ensembles reaches equilibrium in limited regions of the parameter space of external intensive and external extensive variables. In the other regions of the parameter space, the system spontaneously diverges and stays in non-equilibrium although thermodynamic degrees of freedom are satisfied. Here we reveal and determine the divergence threshold and equilibrium conditions of the binary systems. This divergence is found uniquely in the binary and multicomponent systems.

\begin{figure*}[!htb]
	\centering
	\includegraphics[clip]{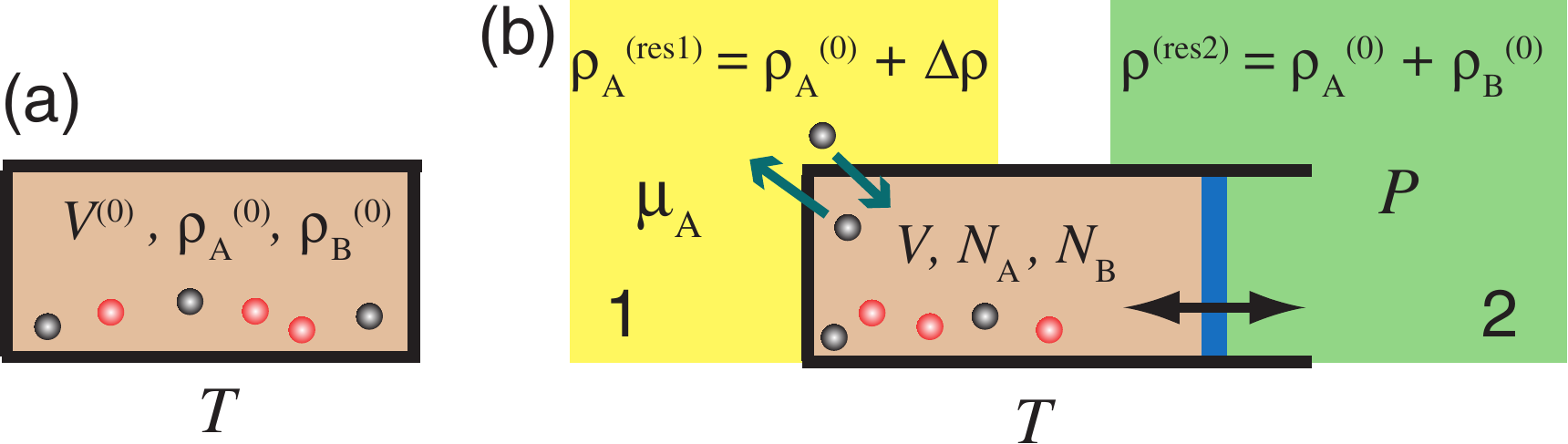}
	\caption{(a) A sketch of a binary ideal gas system at $\left( T, V^{(0)}, N_\text{A}^{(0)}, N_\text{B}^{(0)} \right)$ in the canonical ensemble ($N_\text{A} N_\text{B} VT$-ensemble). $\rho_\alpha^{(0)} = \left. N_\alpha^{(0)} \middle/ V^{(0)} \right.$. (b) A sketch of a binary ideal gas in $\mu_\text{A} N_\text{B} PT$-ensemble. The reservoir 1 and the system are allowed to exchange A-particles to fix the chemical potential of A-species of the system at $\mu_\text{A}$. $N_\text{A}$ of the system is a dynamic variable while $N_\text{B} = N_\text{B}^{(0)}$ is fixed. A piston, denoted by blue, is inserted between the reservoir 2 and the system. The piston adjusts the volume $V$ to fix the pressure of the system at $P$. The system and these reservoirs are connected to a thermostat at $T$, the reservoir 3.
	}
	\label{fig:SketchNANBVTBinarymuANBPTBinaryIdeal}
\end{figure*}
As the simplest example of the binary thermodynamic systems, here we consider a binary ideal gas composed of A and B-species at $\left( T, V^{(0)}, N_\text{A}^{(0)}, N_\text{B}^{(0)} \right)$ in the canonical ensemble ($N_\text{A} N_\text{B} VT$-ensemble), where the index ``${}^{(0)}$'' represents the system in the canonical ensemble. This system is sketched in Fig.~\ref{fig:SketchNANBVTBinarymuANBPTBinaryIdeal}(a). The total pressure of this system is denoted by $P^{(0)} = \left( \rho_\text{A}^{(0)} + \rho_\text{B}^{(0)} \right) k_B T$, where $\rho_\alpha^{(0)} = \left. N_\alpha^{(0)} \middle/ V^{(0)} \right.$ and $k_B T$ is the thermal energy. $\mu_\text{A}^{(0)}$ is the chemical potential of A-species of this system in $N_\text{A} N_\text{B} VT$-ensemble. Connecting reservoir 1, i.e. a reservoir of the chemical potential of A-species, and reservoir 2, a reservoir of the total pressure, to this system, we switch the ensemble of this system. This is sketched in Fig.~\ref{fig:SketchNANBVTBinarymuANBPTBinaryIdeal}(b). The reservoirs 1 and 2 define the values of $\mu_\text{A}$ and total pressure $P$ of the system respectively. The number of B-particles $N_\text{B} = N_\text{B}^{(0)}$ is fixed. These reservoirs and system are connected to the same thermostat at $T$, reservoir 3. The ensemble of this system is classified as one of the conventional ensembles because three external intensive and one external extensive variables, i.e. $( T, P, \mu_\text{A}, N_\text{B} )$, are given from the outside of the system ($\mu_\text{A} N_\text{B} PT$-ensemble). The system volume and number of A-particles of the system are spontaneously determined and denoted by $V$ and $N_\text{A}$ respectively. $\rho_{\alpha} = N_{\alpha} / V$ denotes the number density of $\alpha$-particles of the system. The partial pressure of $\alpha$-species of the system is denoted by $P_\alpha = \rho_{\alpha} k_B T$. $P = \sum_{\alpha} P_\alpha = \rho k_B T$ denotes the total pressure of the system where $\rho = \sum_{\alpha} \rho_{\alpha}$ is the total number density of the particles of the system.

We assume that reservoir 1 is composed of A-particles and contains no B-particles. The number density of A-particles of reservoir 1, denoted by $\rho_\text{A}^\text{(res1)} = \rho_\text{A}^{(0)} + \varDelta \rho$ where $\varDelta \rho > 0$, determines $\mu_\text{A}$. This positive $\varDelta \rho$ indicates $\mu_\text{A} > \mu_\text{A}^{(0)}$. Therefore, according to the equilibrium condition of the chemical potential between the system and reservoir 1, reservoir 1 continues transferring A-particles to the system and increasing $\rho_\text{A}$ until $\rho_\text{A}$ and $\rho_\text{A}^\text{(res1)}$ are consistent. This raises $P_\text{A}$.

Reservoir 2 is composed of both A and B-particles with the total number density $\rho^\text{(res2)} = \rho_\text{A}^{(0)} + \rho_\text{B}^{(0)}$, which results in the total pressure of this resvoir 2 equal to $P^{(0)}$. When reservoir 1 increases $\rho_\text{A}$ and $P = \sum_{\alpha} P_\alpha$, according to the equilibrium condition of the total pressure between the system and reservoir 2, reservoir 2 raises $V$ aiming for small values of $\rho$ and $P$ until the conditions $\rho = \rho^\text{(res2)} = \rho_\text{A}^{(0)} + \rho_\text{B}^{(0)}$ and $P = P^{(0)} = \left( \rho_\text{A}^{(0)} + \rho_\text{B}^{(0)} \right) k_B T$ are satisfied.

The above equilibrium conditions between the system and reservoirs 1 and 2 in $\mu_\text{A} N_\text{B} PT$-ensemble are,
\begin{align}
	\label{eq:EquilibriumChemicalPotentialBinary}
	\mu_\text{A} / k_B T &= \log \left( \varLambda_\text{A}^3 \rho_\text{A}^\text{(res1)} \right)  \notag \\
	&= \log \left( \varLambda_\text{A}^3 \left( \rho_\text{A}^{(0)} + \varDelta \rho \right) \right) = \log \left( \varLambda_\text{A}^3 N_\text{A} \middle/ V \right),
\end{align}
\begin{align}
	\label{eq:EquilibriumPressureBinary}
	P &= P^{(0)} = \rho^\text{(res2)} k_B T  \notag \\
	&= \left( \rho_\text{A}^{(0)} + \rho_\text{B}^{(0)} \right) k_B T = \left( \left( N_\text{A} + N_\text{B} \right) / V \right) k_B T,
\end{align}
where $\varLambda_\alpha = h \left/ \sqrt{2 \pi m_\alpha k_B T} \right.$ denotes the thermal de Broglie wave length of $\alpha$-species~\cite{Norizoe:2014JPSJ},
$h$ is Planck's constant, and $m_\alpha$ mass of the $\alpha$-particle. These simultaneous eqns.~\eqref{eq:EquilibriumChemicalPotentialBinary} and \eqref{eq:EquilibriumPressureBinary} determine values of $V$ and $N_\text{A}$ in equilibrium in $\mu_\text{A} N_\text{B} PT$-ensemble. These conditions are reduced to the equilibration of the particle densities between the system and the reservoirs,
\begin{gather}
	\label{eq:EquilibriumChemicalPotentialBinaryDensity}
	\rho_\text{A}^\text{(res1)} = \rho_\text{A}^{(0)} + \varDelta \rho = N_\text{A} / V,  \\
	\label{eq:EquilibriumPressureBinaryDensity}
	\rho^\text{(res2)} = \rho_\text{A}^{(0)} + \rho_\text{B}^{(0)} = \left( N_\text{A} + N_\text{B} \right) / V.
\end{gather}
These reduced eqns. are solved for $V$,
\begin{equation}
	\label{eq:EquilibriumVBinary}
	V = N_\text{B} \left/ \left( \rho_\text{B}^{(0)} - \varDelta \rho \right) \right. .
\end{equation}
Equation~\eqref{eq:EquilibriumVBinary} indicates that, when $\varDelta \rho$ and chemical potential of reservoir 1 are raised from $\varDelta \rho = 0$ and reach $\varDelta \rho = \rho_\text{B}^{(0)}$, the system diverges ($V, N_\text{A} \to \infty$). In other words, the system is always in the non-equilibrium state in $\varDelta \rho \ge \rho_\text{B}^{(0)}$ although the system stays in equilibrium and keeps finite values of both $V$ and $N_\text{A}$ in $\varDelta \rho < \rho_\text{B}^{(0)}$. This $\varDelta \rho = \rho_\text{B}^{(0)}$ serves as the divergence threshold.

Illustrating the phase behavior at the divergence threshold, $\varDelta \rho = \rho_\text{B}^{(0)}$, here we intuitively demonstrate the divergence of the binary ideal gas system in $\mu_\text{A} N_\text{B} PT$-ensemble at the threshold, $\rho_\text{A}^\text{(res1)} = \rho_\text{A}^{(0)} + \varDelta \rho = \rho_\text{A}^{(0)} + \rho_\text{B}^{(0)}$. The system and reservoir 1 continue to exchange A-particles independently of the value sets of $\rho_\text{B}$ and $V$ until the equilibrium condition between the system and reservoir 1,
\begin{equation}
	\label{eq:EquilibriumChemicalPotentialBinaryDensityThreshold}
	\rho_\text{A}^\text{(res1)} = \rho_\text{A} = \rho_\text{A}^{(0)} + \rho_\text{B}^{(0)},
\end{equation}
which is consistent with eq.~\eqref{eq:EquilibriumChemicalPotentialBinaryDensity}, is fulfilled. On the other hand, this $\rho_\text{A}^\text{(res1)}$ equals the total density of reservoir 2, i.e. $\rho_\text{A}^\text{(res1)} = \rho_\text{A}^{(0)} + \rho_\text{B}^{(0)} = \rho^\text{(res2)}$. Therefore, when eq.~\eqref{eq:EquilibriumChemicalPotentialBinaryDensityThreshold} is fulfilled, the partial pressure of A-species of the system, $P_\text{A}$, is consistent with the total pressure of reservoir 2. If all the B-particles are removed from the system box ($N_\text{B}, \rho_\text{B} \to 0$), this condition simultaneously satisfies the equilibration between the system and reservoirs 1 and 2, which corresponds to the thermodynamically stable point of the single component $\mu PT$-ensemble~\cite{Norizoe:2014JPSJ} of A-species, and results in the equilibrium state in which any intensive variables of the system are determined and any extensive variables always fluctuating and indeterminate. Here, the single component $\mu PT$-ensemble, which is a non-conventional ensemble, denotes the ensemble of single component systems simultaneously connected to 3 reservoirs of the three intensive variables $T$, $P$, and $\mu$. Although, in experiment, arbitrary value sets of these three intensive variables are given from the outside independently from eq.~\eqref{eq:Gibbs-DuhemMulti}, the thermodynamically stable points are defined as points in the 3-dimensional parameter space of $( T, P, \mu )$ where eq.~\eqref{eq:Gibbs-DuhemMulti} is satisfied. However, for $N_\text{B} = N_\text{B}^{(0)} > 0$,
\begin{equation}
	\label{eq:DiffInTotalDensityBetweenSysAndRes2}
	\rho = \rho_\text{A} + \rho_\text{B} = \rho_\text{A}^{(0)} + \rho_\text{B}^{(0)} + \rho_\text{B} = \rho^\text{(res2)} + \rho_\text{B} > \rho^\text{(res2)}.
\end{equation}
Equation~\eqref{eq:DiffInTotalDensityBetweenSysAndRes2} indicates that the total pressure of the system exceeds the total pressure of reservoir 2 by the partial pressure of B-species in the system box, $P_\text{B}$, whenever the equilibrium condition between the system and reservoir 1, eq.~\eqref{eq:EquilibriumChemicalPotentialBinaryDensityThreshold}, is satisfied. This difference in the total pressure between the system and reservoir 2 increases $V$ and decreases $\rho_\text{A} = N_\text{A} / V$, which subsequently raises $N_\text{A}$ and $\rho_\text{A}$ again according to eq.~\eqref{eq:EquilibriumChemicalPotentialBinaryDensityThreshold}. This infinite cycle results in $V, N_\text{A} \to \infty$. In other words, the system diverges for the sake of eliminating $\rho_\text{B} = N_\text{B} / V$ and $P_\text{B}$, i.e. $\rho_\text{B}, P_\text{B} \to 0$. Therefore, the speed of the divergence decreases to zero as $V$ rises and $\rho_\text{B}$ decreases.

On the other hand, for the case $\varDelta \rho > \rho_\text{B}^{(0)}$, when eq.~\eqref{eq:EquilibriumChemicalPotentialBinaryDensity} is fulfilled, the partial pressure $P_\text{A}$ exceeds the total pressure of reservoir 2 even if $N_\text{B}, \rho_\text{B} \to 0$. This condition is inconsistent with the thermodynamically stable point of the single component $\mu PT$-ensemble of A-species even if $N_\text{B}, \rho_\text{B} \to 0$ and results in the divergence of the system for arbitrary values of $N_\text{B}$. Finite $N_\text{B} = N_\text{B}^{(0)} > 0$ raises the total pressure of the system and speed of the divergence. This increase in the speed decreases to zero with $V$.

In $\varDelta \rho < \rho_\text{B}^{(0)}$, if $N_\text{B}, \rho_\text{B} \to 0$, the system vanishes ($V, N_\text{A} \to 0$) because the condition is inconsistent with the thermodynamically stable point of the single component $\mu PT$-ensemble of A-species. However, the finite and fixed $N_\text{B}$ and concomitant $P_\text{B}$ of the present system hold the piston. The system reaches the equilibrium state with the system volume $V$ given in eq.~\eqref{eq:EquilibriumVBinary}.

The divergence of the system in $\mu_\text{A} N_\text{B} PT$-ensemble is similar to the divergence of the single component systems simultaneously connected to the 3 reservoirs $\mu$, $P$, and $T$~\cite{Norizoe:2014JPSJ}. However, the system in $\mu_\text{A} N_\text{B} PT$-ensemble does not vanish unlike such single component systems. Moreover, such single component systems diverge or vanish only when the given value set of external intensive variables $(T, P, \mu)$ violates Gibbs-Duhem equation and, in other words, thermodynamic degrees of freedom. On the other hand, the present binary system diverges although the thermodynamic degrees of freedom, which equal 3, are satisfied. Therefore, the divergence of the system in $\mu_\text{A} N_\text{B} PT$-ensemble is distinct from the divergence of the single component system in $\mu PT$-ensemble and is a phenomenon unique to the binary systems in conventional ensembles.

We have studied and revealed physical properties of the binary $\mu_\text{A} N_\text{B} PT$-ensemble based on the phase behavior of the single-component $\mu PT$-ensemble in the present work. On the other hand, early researchers~\cite{Guggenheim:1939,Prigogine:1950,Hill:StatisticalMechanicsPrinciplesAndSelectedApplications,Sack:1959} also discussed the ensemble of single-component systems connected to the three reservoirs and the ensemble of binary systems connected to the reservoirs of $\mu_\text{A}, P$, and $T$. Their ensembles are similar to, but distinct from, our $\mu PT$-ensemble and $\mu_\text{A} N_\text{B} PT$-ensemble, respectively~\cite{Norizoe:2014JPSJ}. Differences between the present and early works and problems of the early works are illustrated in the following.

Statistical weight (Boltzmann factor) of the ensemble of the single-component system with the three reservoirs was formally introduced by Guggenheim~\cite{Guggenheim:1939}. This ensemble was also studied by other early researchers, \textit{e.g.} Prigogine~\cite{Prigogine:1950} and Hill~\cite{Hill:StatisticalMechanicsPrinciplesAndSelectedApplications}, later. Aiming at revealing a generalized and universal expression for partition functions that are applicable to arbitrary thermodynamically acceptable ensembles~\cite{Sack:1959,Koper:1996}, the early researchers focused on formulation and formalism of the single-component ensemble with the three reservoirs.

Guggenheim assumed that the thermal averages of arbitrary extensive variables, e.g. the system volume and number of particles, were well-defined and finite in the ensemble of single-component systems simultaneously connected to the three reservoirs, when the Boltzmann factor of this ensemble was introduced based on analogy between other already-known conventional ensembles~\cite{Guggenheim:1939,Koper:1996}. However, as was noted by Prigogine and Sack~\cite{Prigogine:1950,Sack:1959}, this assumption contradicts the indetermination of the extensive variables. Demonstrating that the summation of the resulting statistical weight over the phase space, which corresponds to the partition function, diverges, Prigogine concluded that the partition function of this ensemble with the three reservoirs contained no physical meanings~\cite{Prigogine:1950,Sack:1959}. Prigogine's criticism also indicates that the thermodynamic potential of this ensemble, which is defined based on such partition function, would be indefinite. The answer to Prigogine's criticism remained open in the early works since the initial Guggenheim's work.

These problems also apply to the other works which were made based on the Guggenheim's article. For example, the statistical mechanics of the ensemble with the three reservoirs was studied in Hill's book~\cite{Hill:StatisticalMechanicsPrinciplesAndSelectedApplications} based on and according to the above Guggenheim's work and assumption~\cite{Guggenheim:1939}. However, the thermal averages of the extensive variables obviously diverge and are not well-defined in this ensemble. Furthermore, artificial upper cut-off was set to the extensive variables in this Hill's book, based on the assumption of the well-defined and finite thermal averages. These artificial upper cut-off values suppress the divergence of the partition function, whereas artefacts due to this assumption should be introduced to the physical properties of the ensemble with the three reservoirs. Applying the same formulation, Hill also formally derived the statistical weight of the binary system simultaneously connected to the reservoirs of $\mu_\text{A}, P$, and $T$ in the same book~\cite{Hill:StatisticalMechanicsPrinciplesAndSelectedApplications}.

No upper cut-off for the extensive variables is assumed in our present and recent works, by contrast to the above Hill's theory. We considered the thermal averages of \textit{densities} of the extensive variables instead of the extensive variables themselves in our recent work~\cite{Norizoe:2014JPSJ}. By this treatment, the existence of the equilibrium probability distribution of our $\mu PT$-ensemble is guaranteed. This indicates that our derivation of our $\mu PT$-ensemble is distinct conceptually from the early researcher's, e.g. Hill's and Guggenheim's, though the final result seems very similar. Although the extensive variables fluctuate in the vicinity of the \textit{a priori} given thermal averages in Hill's ensemble, the extensive variables spontaneously and unboundedly change in our $\mu PT$-ensemble, keeping the values of the intensive variables fixed. These problems also affect the other works which are based on these early works.

Moreover, in contrast to such mathematical aspects of the partition function, Hill totally reserved physical aspects of the ensemble with the three reservoirs for future works. For example, thermodynamics and statistical mechanics outside the thermodynamically stable point in the single-component ensemble with the three reservoirs were reserved in the Hill's book. Hill overlooked the fact that the system always vanishes or diverges outside the thermodynamically stable point. In our recent work, unlike the early researchers, we revealed the physical aspects of our $\mu PT$-ensemble. Other differences and problems of the early works are also summarized in a recent work by Koper and Reiss~\cite{Koper:1996} as well as ours~\cite{Norizoe:2014JPSJ}.

Thus, the above early researchers concentrated on the mathematical aspects of the partition functions of both the ensemble of single-component systems connected to the three reservoirs and ensemble of binary systems connected to the reservoirs of $\mu_\text{A}, P$, and $T$, i.e. formulation and formalism of these two ensembles. Physical aspects of the ensembles have not seriously been studied. In our recent and present works, we have shed light on these problems and revealed physical characteristics of $\mu PT$ and $\mu_\text{A} N_\text{B} PT$-ensembles. This demonstrates that we have found very simple and basic, but never given a thought before, results in the present work.

In conclusion, we have studied thermodynamics of conventional ensembles of binary systems simultaneously connected to reservoirs of chemical potential, pressure, and temperature. The stability and equilibrium conditions of the binary thermodynamic systems in conventional ensembles have been revealed. In single component systems in conventional ensembles, arbitrary value sets of two intensive and one extensive variables can independently and simultaneously be given from the outside. The single component system reaches equilibrium in the whole parameter space of these given intensive and extensive variables. The straightforward extension of this result to the binary systems in conventional ensembles, which suggests that the binary system in conventional ensembles reaches equilibrium in the whole parameter space of the 3 external intensive and one external extensive variables, has been accepted at least since Gibbs-Duhem equation was found. However, we have demonstrated that the binary system in conventional ensembles reaches equilibrium in limited regions of the parameter space. In the other regions, the system stays in non-equilibrium and diverges although any binary conventional ensemble satisfies thermodynamic degrees of freedom, which equal 3. We have revealed these divergence threshold and equilibrium conditions of the binary systems. This divergence is a phenomenon unique to the binary and multicomponent systems. The single component system diverges or vanishes only when the thermodynamic degrees of freedom are violated in $\mu PT$-ensemble. Moreover, the binary system does not vanish in the conventional ensembles because the external extensive variable is fixed and keeps the finite value. As an example of these results, the thermodynamic stability of the binary ideal gases has been exactly determined, which corroborates our findings.
We can straightforwardly confirm that our results also hold for and apply to arbitrary $M (> 1)$-component systems as well as binary systems.

The divergence threshold divides the non-equilibrium and equilibrium states of the multicomponent thermodynamic system. Changing the values of thermodynamic variables given from the outside, i.e. input parameters, one can switch these states. This switch requires no changes of the ensembles of the system. In terms of experiments, fixing and keeping the experimental device configurations and conditions, and controlling the values of the input parameters, one can freely transfer between the two states of the thermodynamic system. This phenomenon is found outside the framework of both equilibrium thermodynamics and equilibrium statistical mechanics, such as conventional theory of phase transition between equilibrium phases, since one of the two states is essentially in non-equilibrium. Therefore, non-equilibrium thermodynamics and non-equilibrium statistical mechanics, which are developing scientific fields, are required if one wishes to further understand this transfer, e.g. in a manner similar to the conventional theory of phase transition. These developing scientific fields could enhance the understanding of the phenomena, such as further dynamic properties of the thermodynamic system.




%

\end{document}